\documentclass[pra,aps,showpacs,twocolumn]{revtex4-1}
\usepackage{dcolumn}
\usepackage{bm}
\usepackage{amsmath,amssymb,amsfonts,latexsym,fancyhdr,graphicx,epstopdf,times,txfonts}

\begin{document}
\title{Effect of Temperature on Non-Markovian Dynamics in Coulomb Crystals}
\author{Massimo Borrelli$^1$ and Sabrina Maniscalco$^{1,2}$}
\affiliation{$^1$ CM-DTC, School of Engineering \& Physical Sciences, Heriot-Watt University, Edinburgh EH14 4AS, United Kingdom\\
$^2$Turku Centre for Quantum Physics, Department of Physics and Astronomy, University of
Turku, FI-20014 Turun yliopisto, Finland}

\begin{abstract}
In this paper we generalize the results reported in {\it Phys. Rev. A} {\bf 88} 010101 (2013) and investigate 
the flow of information induced in a Coulomb crystal in presence of thermal noise. For several temperatures we calculate the non-Markovian character
of Ramsey interferometry of a single $1/2$ spin with the motional degrees of freedom of the whole chain.
These results give a more realistic picture of the interplay between temperature, non-Markovianity and criticality. 
\end{abstract}

\maketitle

\section{Introduction}	
The investigation of collective phenomena within an open quantum system framework\cite{breuerbook} has recently gained
considerable attention. Examples of such studies span from cold atoms\cite{pinja1}, to 
magnetic systems\cite{pinja2,mauro} and fermionic gases\cite{john}. Generally speaking, this novel approach relies on studying some
global properties of a many-body system by probing it with a single and well controllable quantum system.
In this respect, the many-body system is to be seen as a complex environment interacting with an open but yet
well defined quantum system. However, unlike in standard open quantum system theory, the attention here is shifted to the environment. 
The key idea is that by monitoring the dynamics of the open quantum system we gain information about the environment. 
In particular, the loss or temporary re-gain of the system's quantum properties due to the interaction with its surrounding 
environment characterizes the nature of the dynamical process undergone by the system and this is the aspect 
we are most interested in.\\ 
The dynamics of an open quantum system can be either Markovian or non-Markovian.
The Markovian regime is characterized by a complete loss of all the quantum properties. The theory of Markovian
systems is mathematically well formulated and understood: any process of this sort can be described by a completely positive and trace-preserving (CPT) map
that is the solution of a Lindblad master equation for the dynamics of the reduced quantum system only\cite{lindblad}. For non-Markovian systems a number of non-Markovianity measures have been proposed. When the measure is nonzero, then the system is said to be non-Markovian. However, in general these definitions do not coincide.\cite{breuer,plenio,plastina,bogna,sabrina,luo}\\
In this manuscript we extend the investigation performed in Ref. \cite{massimo}. We consider Ramsey interferometry of
a single two-level ion embedded in a Coulomb crystal\cite{gabriele1} where thermal excitations are initially present. A Coulomb crystal is 
an arrangement of magnetically trapped ions that can exhibit multiple spatial configurations depending on the values of some
trapping parameters\cite{morigi1,morigi2,dubin}. Switching from a configuration to another is accompanied by a structural phase transition. In Ref. \cite{massimo}
the non-Markovianity character of the interferometric protocol was studied as a function of the chain's closeness to criticality. To quantify
such a feature we used the non-Markovianity measure introduced in Ref. \cite{breuer}. A significant sensitivity
of the latter quantity to the crossing of the critical point was found along with a recipe to experimentally test the theory. However, the whole
study was carried out at zero temperature. Here, we include the effect of an initially thermally excited environment in order
to provide a more realistic picture. This paper is structured as follows: in section 2 we introduce some basic theory of Coulomb crystals, section
3 illustrates the interferometric protocol studied. In section 4 we briefly we introduce the non-Markovianity measure used in this investigation and in section
five we report the main findings.

\section{Coulomb Crystals}
A Coulomb crystal is a spatial arrangement of $N$ ions of charge $Q$ and mass $m$ where the balance between the ion-ion Coulomb repulsion 
and the confining potential generated by the linear trap results in a regular geometric pattern. The Hamiltonian governing
this system reads as follows\cite{fishman}
\begin{equation}
H=\sum_{j=1}^{N}\frac{p_{j}^{2}}{2m}+\frac{1}{2}m\left[\nu^{2}x_{j}^{2}+\nu_{t}^{2}\left(y_{j}^{2}+z_{j}^{2}\right)\right]+
\frac{1}{2}\sum_{j\ne i=1}^{N}\frac{Q^{2}}{\left|\vec{r}_{j}-\vec{r}_{i}\right|}
\label{hamiltonian}
\end{equation}
where $\vec{p}_{j}$ is the momentum of the $j-$th ion, $\vec{r}_{j}=\left(x_{j},y_{j},z_{j}\right)$ its position, $\nu$ and $\nu_{t}$
the axial and transverse trap frequencies, respectively. Several patterns for the equilibrium configuration can be 
explored by tuning the frequencies appropriately\cite{dubin1}. Furthermore, any drastic change in these patterns is accompanied by a structural phase transition that
can be observed when we modify these parameters continuously and not too abruptly. 
Here, we focus our attention on the first in the hierarchy of these subsequent phase transitions: the linear-to-zig-zag.
As shown in Ref. \cite{fishman}, as long as the transverse frequency fulfills the following condition
\begin{equation}
\nu_{t}>\omega_{0}\sqrt{\frac{7\zeta(3)}{2}}\equiv\nu_{t}^{(c)}
\label{criticality}
\end{equation}
where $\omega_{0}=\sqrt{Q^{2}/ma^{3}}$, the equilibrium configuration is a linear chain where all the ions are equally spaced
and their equilibrium positions are located at $\vec{r}_{j}^{(0)}=\left(ja,0,0\right)$ with $j=1,\dots,N$. This pattern is structurally
equivalent to a simple one-dimensional solid where all the vibrational degrees of freedom are uncoupled from each other.
The normal mode wave vectors of this chain are $k=2\pi n/Na$ where $n=0,1,\dots,N/2$ and the dispersion relations read as\cite{gabriele1} 
\begin{equation}
\begin{aligned}
&\omega_{\parallel}(k)=\omega_{0}\sqrt{8\sum_{j=1}^{N/2}\frac{1}{j^{3}}\sin^{2}\left(\frac{jka}{2}\right)}\\
&\omega_{\perp}(k)=\sqrt{\nu_{t}^{2}-4\omega_{0}^{2}\sum_{j=1}^{N/2}\frac{1}{j^{3}}\sin^{2}\left(\frac{jka}{2}\right)}
\end{aligned}
\label{dispersion}
\end{equation}
If, on the contrary, Eq.\eqref{criticality} is not fulfilled, the ions reorganize in a two-dimensional zig-zag configuration with the new
equilibrium positions being $\vec{r}_{j}^{(0)}=\left(ja,(-1)^{j}b/2,0\right)$. The parameter $b$ is the transverse equilibrium distance, whose value 
can be found in Ref. \cite{fishman}. In this configuration the $x$ and the $y$ modes are coupled to each other and the first Brillouin zone is half the size of the 
linear counterpart, spanning from $0$ to $\pi/2a$ with the normal mode wave vectors being $k=2\pi n/Na$.
The excitation spectrum splits in four branches for each $k$ and it is way more structured than in the simple linear case. 
When tuning the transverse frequency below the critical value $\nu_{t}^{(c)}$ the chain undergoes a second-order structural phase
transition that is driven by the smallest wavelength mode at $k=\pi/a$, whose transverse energy vanishes exactly at criticality. All of the motional eigenmodes, both in the linear
and in the zig-zag regime, can be quantized through standard quantization procedure. Moreover, by means of Taylor expansion up to 
the second order in terms of the ion's displacement from the equilibrium positions, the Hamiltonian \eqref{hamiltonian} can be mapped onto an effective harmonic oscillator. Thus the eigenmodes, within the validity of this approximation, follow a bosonic statistics.   

\section{Ramsey Interferometry in Coulomb Crystals}
As suggested in Ref. \cite{gabriele1} Ramsey interferometry of a single spin $1/2$ can be used to study some collective properties of the chain when
driven across the critical point. In this scheme, two electronic levels of one of the ions in the chain (the target ion) are selected and coupled to the motional 
degrees of freedom of the whole chain via laser pulses of fixed duration. The protocol goes as follows. We label the internal levels of the target ion
$\{|e\rangle,|g\rangle\}$ and define the usual spin $1/2$ relevant operators, $\sigma_{z}=\left(|e\rangle\langle e|-|g\rangle\langle g|\right),\sigma_{+}=|e\rangle\langle g|, \sigma_{-}=|g\rangle\langle e|$. We label the energy separation between these two levels $\hbar\bar{\omega}$. After properly initializing the  two-level system to a well defined initial state $|\phi_{0}\rangle$ we implement the following
laser-assisted interaction
\begin{equation}
H_{\textrm{INT}}=\hbar\Omega\left[\sigma_{+}e^{-i(\omega_{L}t-k_{L}y_{1})}+\textrm{h.c.}\right],
\label{laserinteraction}
\end{equation}
where $\Omega$ is the Rabi frequency of the laser, $\omega_{L}, k_{L}$ are the laser frequency and wave-vector respectively,
and $y_{1}$ is the position operator of the target ion. The position operator $y_{1}$ can be expanded in terms of the raising and lowering operators of the normal modes
and, when exponentiated, it generates a multi-mode displacement operator for the transverse eigenmodes of the chain\cite{gabriele1}
\begin{equation}
e^{-ik_{L}y_{1}}=\bigotimes_{j}D(\alpha_{j})=\bigotimes_{j}\exp\left(\alpha_{j}b^{\dagger}_{j}-\alpha^{*}_{j}b_{j}\right),
\label{displ}
\end{equation}
where $b_{j},b^{\dagger}_{j}$ are the ladder operators of the $j-$th transverse mode such that $[b_{j},b^{\dagger}_{k}]=\delta_{jk}$,  $\alpha_{j}=i\eta\sqrt{\omega_{0}/2\omega_{j}}S_{1j}$, the Lamb-Dicke parameter is $\eta=k_{L}\sqrt{\hbar/m\omega_{0}}$ and the matrix $S_{ij}$, which realizes the normal modes decomposition, is defined in Ref. \cite{gabriele1}. The $j$ index encodes all the quantum numbers defining the modes of the environment (momentum and parity). Hence, the target ion receives a state-dependent 'kick' in the transverse $y$ direction, starts Rabi-oscillating between the two internal states and excites
all the transverse vibrational normal modes of the chain.
After this first pulse both the two-level system and the whole chain
are let to relax and evolve freely according to the decoupled Hamiltonian 
\begin{equation}
H_{0}=\hbar\bar{\omega}/2\sigma_{z}+\sum_{j}\hbar\omega_{j}b_{j}^{\dagger}b_{j}
\label{freehamiltonian}
\end{equation}
After a fixed time $t$ an opposite laser pulse is applied. By looking at the time-dependent visibility of the Ramsey fringes, which is ultimately connected to the ground state
probability of the single spin, it is possible to study the spatial autocorrelation function of the whole chain. Hence, some of the information regarding
the dynamics of a complex many-mode bosonic system, is mapped onto the dynamics of a single two-level system. In particular, in Ref. \cite{gabriele1} it was found that the closer the chain is to
criticality, the faster the interferometric signal decays. In Ref. \cite{massimo} the authors take a step further in this analysis. The Ramsey interferometric scheme is reinterpreted as an open quantum system process where the single spin is treated as the system and the motional degrees of freedom of the whole chain represent the environment. In this spirit, the natural questions to address regards the nature of this process, Markovian or non-Markovian, and whether a suitably defined measure for the non-Markovian character of a process could be used to gain information about the critical environment. The answer is positive and, whereas the process is non-Markovian as long as the environment is pushed away from criticality, its behavior is largely different when $\nu_{t}\approx\nu_{t}^{(c)}$. In this limit, the Markovian nature of the Ramsey interferometry protocol arises.
This analysis in Ref. \cite{massimo} was carried out in the $T=0$ limit with no excitations initially present in the environment. In the following we shall study how these results are affected when the environment is thermally excited at the beginning of the Ramsey  protocol.

\section{Non-Markovianity Measure}
In this article, in order to witness and quantify the non-Markovian character of a quantum process we make use of the definition first introduced in Ref. \cite{breuer} which relies on the idea of distinguishability of quantum states. Given two generic quantum states $\rho_{1}, \rho_{2}$ we make use of the trace-distance, defined as 
\begin{equation}
D(\rho_{1},\rho_{2})=\frac{\textrm{tr}\left|\rho_{1}-\rho_{2}\right|}{2}
\label{tracedistance}
\end{equation}
as a mathematical tool to distinguish them. When dealing with a Markovian process, any pair of two initially well distinguishable states can only become more and more indistinguishable as time goes by. This translate to a trace distance that cannot ever increase: Markovian processes are contractive
\begin{equation}
D(\rho_{1}(t),\rho_{2}(t))\geq D(\rho_{1}(t+\delta t),\rho_{2}(t+\delta t))\qquad\forall\rho_{1},\rho_{2},\forall t,\delta t
\label{contractivity}
\end{equation}
Thus, in order to witness non-Markovianity in an open quantum system, a violation of condition \eqref{contractivity} for at least one pair of states and one time interval is sufficient. In order to further quantify the degree of non-Markovianity of a process the authors of Ref. \cite{breuer} introduced the following measure 
\begin{equation}
\mathcal{N}(\Lambda)=\max_{\rho_{1,2}(0)}\int_{\sigma>0}dt\sigma(t,\rho_{1,2}(0)),
\label{nm}
\end{equation}
where $\sigma(t,\rho_{1,2}(0))=\frac{d}{dt}D(\rho_{1}(t),\rho_{2}(t))$ can be seen as the rate at which the information regarding the open system is lost or partly re-gained due to memory effects. The maximization is to be performed over all the possible pairs of initial states in the state space of the open system.
Since in the following we are going to be interested in a simple two-level system, we only have to care about pairs of initial pure states when it comes to the maximization procedure\cite{antti}. These pair of states are represented by antipodal points in the Bloch-sphere visualization.

\section{Temperature}
\begin{figure}[h!]
\centering
\includegraphics[scale=0.8]{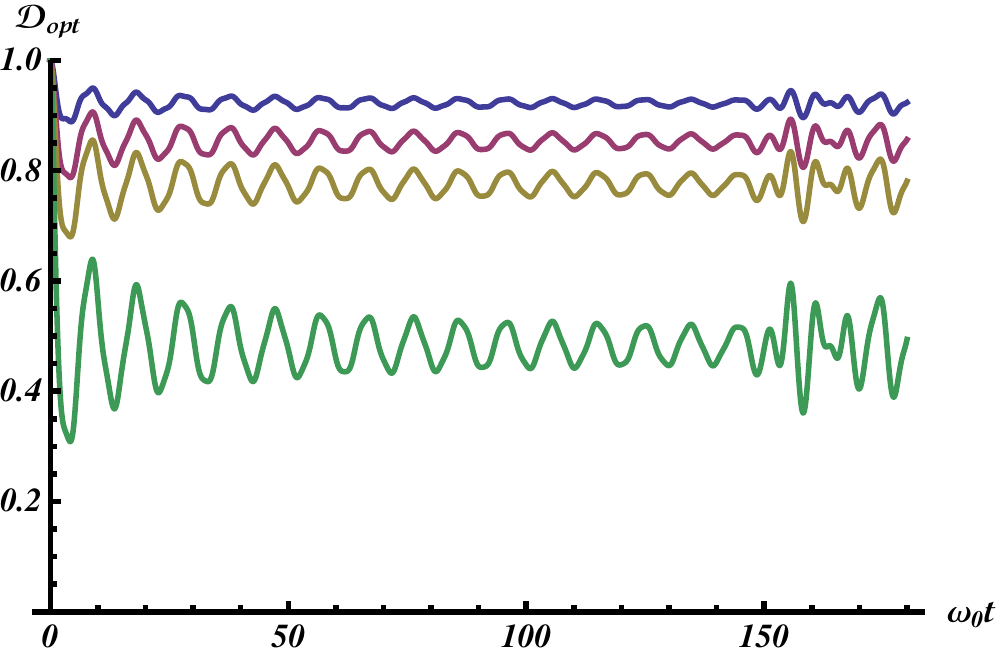} 
\vspace*{8pt}
\caption{Time-evolution of $D_{opt}(t,\beta)$ at $\Delta=0.1$ for $N=100$ and four different values of temperature: $\beta\hbar\omega_{\textrm{max}}=0.3$ purple, $\beta\hbar\omega_{\textrm{max}}=0.7$ dark pink, $\beta\hbar\omega_{\textrm{max}}=1.2$ dark yellow and $\beta\hbar\omega_{\textrm{max}}=4.3$ green.}
\label{plot1}
\end{figure}

\begin{figure}[h!]
\includegraphics[scale=0.8]{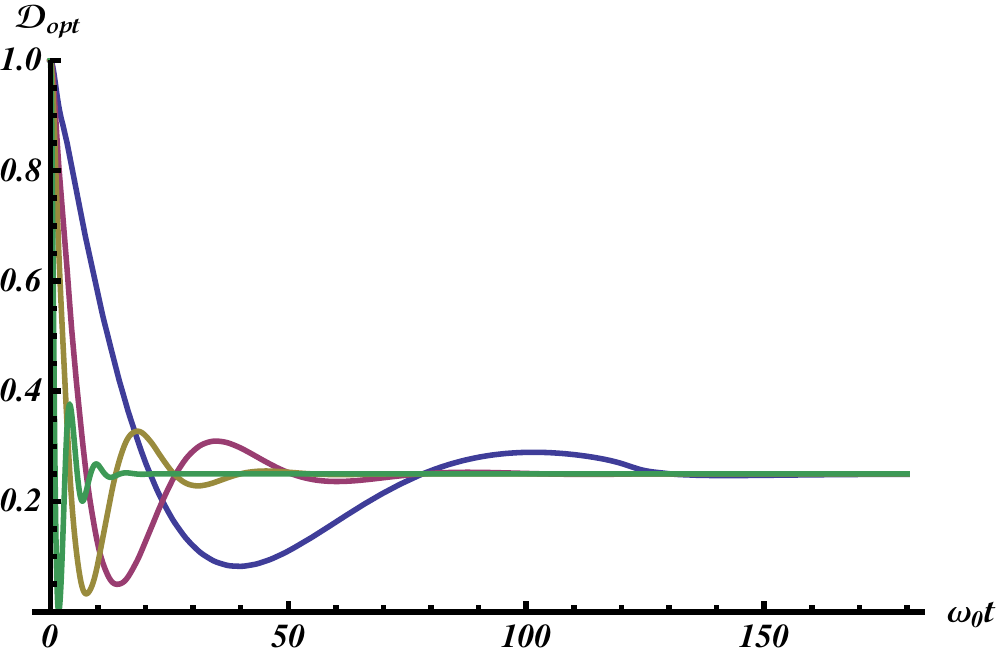}  
\vspace*{8pt}
\caption{Time-evolution of $D_{opt}(t,\beta)$ at $\Delta=10^{-5}$ for $N=100$ and four different values of temperature: $\beta\hbar\omega_{\textrm{max}}=0.3$ purple, $\beta\hbar\omega_{\textrm{max}}=0.7$ dark pink, $\beta\hbar\omega_{\textrm{max}}=1.2$ dark yellow and $\beta\hbar\omega_{\textrm{max}}=4.3$ green.}
\label{plot2}
\end{figure}

\begin{figure}[h!]
\includegraphics[scale=0.8]{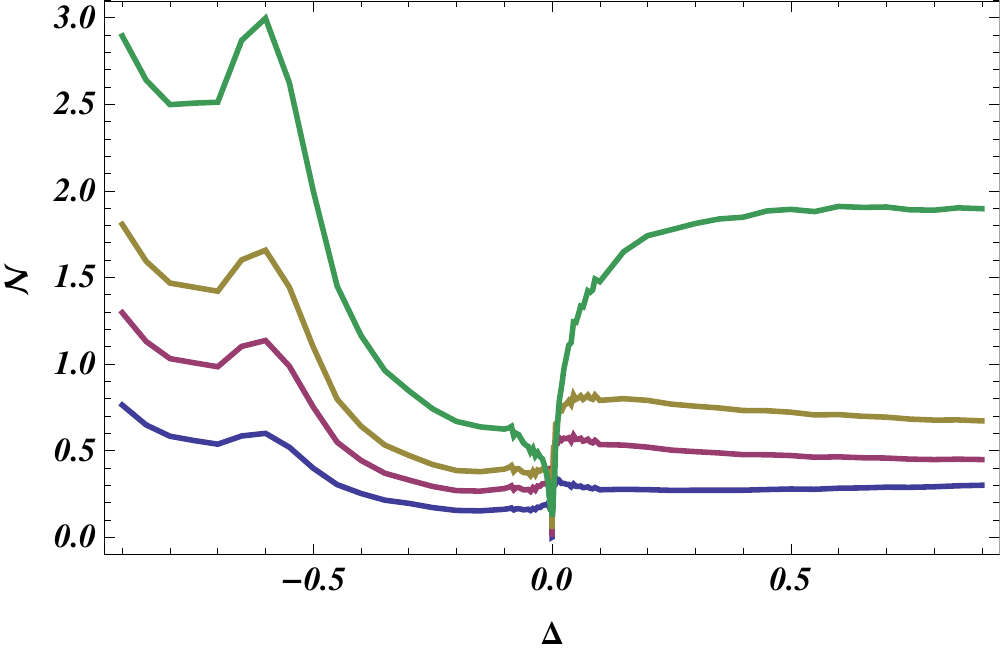}  
\vspace*{8pt}
\caption{Non-Markovianity measure $\mathcal{N}$ as a function of the relative distance $\Delta$ for four different values of temperature: $\beta\hbar\omega_{\textrm{max}}=0.3$ purple, $\beta\hbar\omega_{\textrm{max}}=0.7$ dark pink, $\beta\hbar\omega_{\textrm{max}}=1.2$ dark yellow and $\beta\hbar\omega_{\textrm{max}}=4.3$ green. The truncation time in $\mathcal{N}$ is about $\omega_{0}t_{T}\approx120$.}
\label{plot3}
\end{figure}

In this section we report the main findings of this article. As mentioned above we focus on the case where at the beginning of the Ramsey protocol thermal excitations in the environment are present. This translates to the following initial join state of system and environment
\begin{equation}
\rho_{I}=|\phi_{0}\rangle\langle\phi_{0}|\otimes\rho_{T}
\label{initialstate}
\end{equation}
where the initial system pure state is a generic linear superposition and the environment state is a multi-mode factorized thermal state
\begin{equation}
\begin{aligned}
&|\phi_{0}\rangle=\cos\left(\frac{\theta}{2}\right)|e\rangle+e^{i\phi}\sin\left(\frac{\theta}{2}\right)|g\rangle\\
&\rho_{T}=\bigotimes_{k,\sigma}\left(\sum_{n_{k,\sigma}=0}^{\infty}\frac{e^{-\beta\hbar\omega(k)n_{k,\sigma}}}{Z_{k,\sigma}}|n_{k,\sigma}\rangle\langle n_{k,\sigma}|\right)
\label{states}
\end{aligned}
\end{equation}
where $\beta=1/k_{B}T, \omega(k)\equiv\omega_{\perp}(k)$, $Z_{k,\sigma}=\sum_{n_{k,\sigma}=0}^{\infty}e^{-\beta\hbar\omega(k)n_{k,\sigma}}$ and $\sigma$ indicates the mode parity. The reduced density matrix for the system dynamics can be easily obtained from the global dynamics by use of partial trace. The total evolution operator reads as follows
\begin{equation}
U(t)=U_{INT}(-\pi/2)U_{0}(t)U_{INT}(\pi/2).
\label{utotale}
\end{equation}
where $U_{INT}(\pi/2)$ represents the initial pulse, see Eq. \eqref{laserinteraction}, $U_{0}(t)$ the free evolution dictated by Hamiltonian \eqref{freehamiltonian} and $t$ the time elapsed in between the two opposite pulses. Thus, the system density operator at the end of the whole protocol for a fixed $t$ reads as
\begin{equation}
\rho_{S}(t)=\textrm{tr}_{E}\left[U(t)\rho_{I}U(t)^{\dagger}\right]
\label{reducedsystem}
\end{equation}
Obviously, since both the displacement amplitudes $\alpha_{j}$ and the free Hamiltonian depend upon the value of the transverse frequency, we expect that changes in such a parameter will result in different outcomes at the end of the Ramsey protocol. Hence, we can compute the non-Markovianity measure \eqref{nm} for different values of $\nu_{t}$ and study its behavior close to criticality. Also, we expect the value of $\mathcal{N}$ to vary whenever the upper limit of integration in \eqref{nm} changes.
First we look at the time-evolution of the trace distance. In the $T=0$ limit it was shown numerically that the maximizing pair is formed by the eigenstates of $\sigma_{x}$, which we label $|+\rangle, |-\rangle$\cite{massimo}. Here, we assume that, as long as the temperature of the environment is not to large as compared to the its highest frequency, the maximazing pair in \eqref{nm} is still the same as in the zero-temperature case. The trace distance, in analogy to the $T=0$ case, can be written as
\begin{equation}
D_{opt}(t,\beta)\approx\frac{1}{4}\left|1+2\cos\left[B(t,\beta)\right]\left(\mathcal{V}(t,\beta)-\frac{\xi^{4}(\beta)}{\mathcal{V}(t,\beta)}\right)+\mathcal{V}^{4}(t,\beta)+2\xi^{4}(\beta)\right|,
\label{tracedistopt}
\end{equation}
However, a finite temperature modifies all the following quantities
\begin{equation}
\begin{aligned}
&A(t,\beta)=2\sum_{j}|\alpha_{j}|^{2}\coth\left(\frac{\hbar\omega_{j}\beta}{2}\right)\sin^{2}\left(\omega_{j}t/2\right)\\
&B(t,\beta)=\sum_{j}|\alpha_{j}|^{2}\coth\left(\frac{\hbar\omega_{j}\beta}{2}\right)\sin(\omega_{j}t)\\
&\xi(\beta)=e^{-\sum_{j}|\alpha_{j}|^{2}/2}\coth\left(\frac{\hbar\omega_{j}\beta}{2}\right)\\
&\mathcal{V}(t,\beta)=\exp\left[-A(t,\beta)\right]
\label{quantities}
\end{aligned}
\end{equation}
where the index $j$ again encodes momentum $k$ and parity $\sigma$. We define the parameter $\Delta=\nu_{t}/\nu_{t}^{(c)}-1$ as the relative distance from criticality.
Fig.\eqref{plot1} shows the time-evolution of the optimal trace distance for different values of the initial environment temperature and for  $\Delta=0.1$, thus in the linear configuration, and for $N=100$. Increasing the temperature results in wider oscillations that get eventually damped up to a revival time  $\omega_{0}t_{R}\approx140$. It is worth noticing that in this regime the oscillations are roughly in phase regardless of the environment temperature.  If we push the chain close to criticality we observe a completely different behavior, which is displayed in Fig.\ref{plot2}. In this case, when the temperature of the environment increases, the trace distance's decay is much faster while the amplitude of the few oscillations does not appear to be greatly affected. Also, the saturation value, roughly $0.25$, is independent of the temperature.
\\
In order to have quantitative and time-independent results connecting the flow of information and the criticality of the environment we now study the behavior of $\mathcal{N}$ as a function of $\Delta$, see Fig.\ref{plot3}. We choose the truncation time, that is the upper integration limit, $\omega_{0}t_{T}\approx120$: this guarantees no physical excitation has yet gone back to the system and all the contributions to $\mathcal{N}$ solely come from backflow of information. We notice a clear dip in $\mathcal{N}$ in the proximity of the critical point, located at $\Delta=0$. It is important to remark that, in the actual numerical calculation performed here, the smallest $\Delta$ achievable is $\Delta=10^{-5}$. This limit guarantees that $4^{\textrm{th}}$ order effects can be safely neglected and the harmonic approximation is valid and nicely working. By inspection of Fig.\ref{plot3} we notice that when the temperature of the environment increases the steepness of the dip decreases. This effect is especially obvious when looking at the green curve, corresponding to the highest temperature, on the zig-zag side of the transition. It is also interesting to notice that far from criticality the higher the temperature the more non-Markovian the environment: this is in agreement with the dynamics of the trace distance displayed in Fig.\ref{plot1}. Such a result might initially sound counterintuitive as one would expect an increase in the temperature of the environment to make the system more Markvovian. However, it can be understood with a simple argument. Since the environment is initialized in a thermal state, several modes are already populated prior to the interferometric protocol. The coupling between the target ion and the rest of the chain does not depend on the energy of each mode. Thus, the more modes are initially excited, the more modes the target ion will couple to, regardless of their energy. This reflects in the fact that when we increase the initial temperature the size of the environment, that is the amount of available modes for the system to interact with, increases leading to larger exchange of information.  

\section*{Conclusions}
In this article we extend the model presented in Ref. \cite{massimo} including a finite temperature of the environment. We find that even in this case 
the non-Markovianity measure is extremely sensitive to the phase transition and remarkably pinpoints the critical point. This system is Markovian only at criticality. Furthermore,
increasing the temperature of the environment leads to a larger non-Markovian character of the Ramsey interferometric scheme on both
sides of the phase transtion. These results provide a more realistic description of how the backflow of information behaves in such a system
whenever a residual thermal character, due to a non-perfect state preparation of the environment state, is present prior to the execution of the protocol.

\section*{Acknowledgments}

M.B. would like to thank EPSRC CM-DTC for financial support and Gabriele De Chiara for useful discussion.

\end{document}